# Sound Signal Synthesis with Auxiliary Classifier GAN, COVID-19 cough as an example

Yahya Sherif Solayman Mohamed Saleh[1], Ahmed Mohammed Dabbous[2], Lama Alkhaled[2], Hum Yan Chai[3]*, Muhammad Ehsan Rana1[1], Hamam Mokayed[2]*

[1]Faculty of Computing, Engineering and Technology, Asia Pacific University, Kuala Lumpur, Malaysia.
[2]Department of Computer Science, Electrical and Space Engineering, Lulea University of Technology, Sweden.
[3]Department of Mechatronics and Biomedical Engineering, Lee Kong Chian Faculty of Engineering and Science, Universiti Tunku Abdul Rahman, Malaysia

*Corresponding author: Hamam Mokayed (e-mail: hamam.mokayed@ltu.se).

**Abstract:** One of the fastest-growing domains in AI is healthcare. Given its importance, it has been the interest of many researchers to deploy ML models into the ever-demanding healthcare domain to aid doctors and increase accessibility. Delivering reliable models, however, demands a sizable amount of data, and the recent COVID-19 pandemic served as a reminder of the rampant and scary nature of healthcare that makes training models difficult. To alleviate such scarcity, many published works attempted to synthesize radiological cough data to train better COVID-19 detection models on the respective radiological data. To accommodate the time sensitivity expected during a pandemic, this work focuses on detecting COVID-19 through coughs using synthetic data to improve the accuracy of the classifier. The work begins by training a CNN on a balanced subset of the Coughvid dataset, establishing a baseline classification test accuracy of 72%. The paper demonstrates how an Auxiliary Classification GAN (ACGAN) may be trained to conditionally generate novel synthetic Mel Spectrograms of both healthy and COVID-19 coughs. These coughs are used to augment the training dataset of the CNN classifier, allowing it to reach a new test accuracy of 75%. The work highlights the expected messiness and inconsistency in training and offers insights into detecting and handling such shortcomings.

**Keywords:** Signal processing, sound signals, GAN, COVID-19, cough analysis

## 1. Introduction

Health care is one of the most impactful and important fields in today's industry; it is essential to people's lives and wellness. Yet once again, during the COVID-19 pandemic, the industry's shortcomings were highlighted as the outdated healthcare system struggled to keep up with the ever-growing demand for medical services. As the challenges in health care continue to grow, and as the fast-paced advancements in artificial intelligence continue to take the world by storm, many are exploring the latter to aid in the former. Indeed, health care is one of those vital domains where advancements fall short due to the scarcity of datasets for training. Recording medical data is difficult and expensive, and thus, data is scarce, which limits the effectiveness of deep learning as, by its nature, it requires a huge quantity of data to learn the underlying latent features correctly and show acceptable performance. To alleviate this problem, some researchers turned to deep generative models for data synthesis, where the goal is to leverage a generative model to generate new synthesized medical data that falls within the domain of interest, effectively increasing the size of the dataset for the target domain (Saleh et al., 2023). While common data Augmentation methods exist and are still commonly used (Mokayed et al., 2024), generative models can learn latent features and generate new data with unique features that still fall within the domain of interest, something that data augmentation techniques cannot achieve.

The dominant architecture for this generative task within the context of COVID-19 even in the era of transformers (Vaswani et al., 2017) and diffusion models (Ho et al., 2020) is a Generative Adversarial Network (GAN) (Jiang et al., 2020; Liu et al., 2020; Waheed et al., 2020; Mahapatra & Singh, 2021; Menon et al., 2021; Motamed et al., 2021; Zunair & Hamza, 2021a; Dravid et al ., 2021). A GAN (Goodfellow et al., 2014) is composed of two neural networks designed to be trained in an adversarial matter where one, the generator, is responsible for generating synthetic data that is indistinguishable from the datasets, and the other, the discriminator, is designed to detect generated data and propagate back information to the generator. All the published work in the task of synthesizing for the COVID-19 domain focus on radiological data such as Chest X-rays (CXR) and Computed Tomography (CT) images as a more reliable data form to detect COVID-19 symptoms (Ai et el., 2020; Bernheim et al., 2020; Chung et al., 2020; Fang et al., 2020; Zhao et al., 2020a; Zhao et al., 2020b; Kanne et al., 2020; Guan et al., 2020; Huang et al., 2020) than reverse-transcription polymerase chain reaction (RT-PCR). However, an overlooked method of COVID-19 detection is through cough audio, and while promising work was published showing that deep learning models can detect COVID-19 through coughs (Haritaoglu et al., 2022 Muguli et al., 2021; Orlandic et al., 2021; Södergren et al., 2021), to the best of our knowledge no work was published in the synthesis of COVID-19 coughs. To this end, we propose the use of an Auxiliary classifier GAN (ACGAN) for the generative task of synthesizing COVID-19 positive coughs to proliferate the available datasets in that domain and further the progress of Artificial intelligence in the field of Psychoacoustics.

## 2. Related work

### 2.1 Radiological synthesis of COVID-19 scans

Reviewing all the existing literature in the synthesis of COVID-19 data, we note that it all focused on the radiological domain, synthesizing either a CXR or a CT. The most dominant technique for synthesis was image-to-image translation, where a GAN was used to convert a COVID-19 negative scan into the positive domain. The generated data was then used for dataset expansion and data anonymization (Saleh et al., 2023; Jiang et al., 2020; Waheed et al., 2020; Zunair & Hamza, 2021a; Liu et al., 2020). Additionally, the authors leveraged the GAN's accumulated knowledge for various tasks such as semantic segmentation, direct classification, and saliency map generation.

Semantic segmentation proved to improve the synthesis performance (Motamed et al., 2021) and vice versa (Liu et al., 2020). most works (Motamed et al., 2021; Mahapatra & Singh, 2021; Liu et al., 2020) by leveraging the U-net architecture (Ronneberger et al., 2015). Other works leveraged GANs for direct classification, which demonstrated the GAN's application in semi-supervised learning, which is a robust approach when faced with limited data (Waheed et al., 2020; Motamed et al., 2021). Lastly, GANs were used to generate rich activation maps for COVID-19 scans, an approach that is useful as a sanity check for the model but can also aid in learning and understanding the domain of interest by studying what the features the model focuses on (Dravid & Katsaggelos, 2021; Zunair & Hamza, 2021b).

*2.1.1  COVID-19 detection through coughs*

To the best of our knowledge, there has been no paper published on synthesizing cough audio for COVID-19 detection; however, many published works show the need for such research. In the wake of the COVID-19 pandemic, many researchers explored the role machine learning can play in providing efficient screening. Many of those researchers introduced deep learning-based solutions for detecting the virus through a person's cough. The work by Haritaoglu et al. (2022) tested multiple COVID-19 classification models that work on cough audio. The clear conclusion from evaluating the classification models is that data scarcity leads to a bias in the performance against COVID-19 detection. Another work by Kumar and Pja (2021) applies a one-dimensional CNN to the same problem. They also collected five thousand cough audios to improve the model's performance; however, only three hundred of those were COVID-19 positive. Looking at those findings, there is a huge scarcity of data within this domain, and many entities have taken action to mitigate this limitation. One such entity is the one behind the DiCOVA challenge (Muguli et al., 2021) and (Sharma et al., 2021). The challenge hoped to get more researchers to study sound signals for the task of COVID-19 classification; however, the dataset provided was lackluster and biased towards the negative samples, with only 172 positive cases out of the 965 provided. The efforts in providing suitable datasets for COVID-19 screening through audio to the community continued, most notably the work by Orlandic et al. (2021), who proposed the COUGHVID crowdsourcing dataset, a large-scale dataset for cough analysis. This dataset consists of 20,000 samples with 1,010 being samples of people "claiming to have COVID-19" (Orlandic et al., 2021). The increased dataset size will certainly improve the performance of the models; however, the bias against COVID-19 patients still resides. The first explored work by Haritaoglu et al. (2022) used the COUGHVID along with the Coswara dataset (Sharma et al., 2020) and still reported bias in the model's performance.

## 3. Auxiliary Classifier GAN (ACGAN)

Given the scarcity of audio-positive coughs, we propose the use of an ACGAN to leverage the information in negative data to produce synthetic COVID-19 coughs. ACGANs (Odena et al., 2017) are conditional GANs that condition the generator on the class label by introducing it to both the generator and discriminator, as shown in Figure 1. The generator, G, takes as input a class label c that the generated sample should belong to, $x_g = G(c, z)$. The discriminator is trained to output two probability distributions: over the source, i.e., whether x is real or fake, $y = p(s \vee x)$, and over the class labels $Q = p(c \vee x)$. The generator and discriminator losses are modified to maximize the log-likelihood of the correct class in addition to their original loss, as shown in Equation (1).

$L_D^{GAN} = \max_s \left[ E_{x_r p_r(x)}[log p(s = 1|x_r)] + E_{x_g p_g(x)}\left[ log\left(1 - p(s = 1|x_g)\right)\right]\right] + \max_C \left[ E_{x p_r(x)}[log p(C = c \vee x_r)] + E_{x_g p_g(x)}[log p(C = c \vee x_r)]\right]$

$L_G^{GAN} = \min_G E_{x_g p_g(x)}\left[ log\left(1 - p(s = 1|x_g)\right)\right] + \max_C \left[ E_{x p_r(x)}[log p(C = c \vee x_r)] + E_{x_g p_g(x)}[log p(C = c \vee x_r)]\right]$  (1)

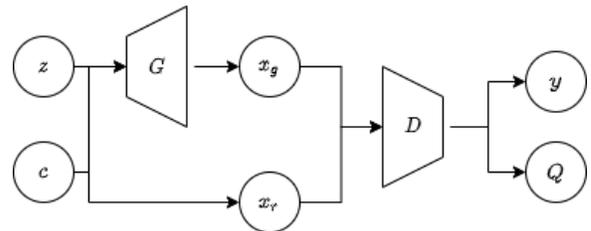

**Figure 1**. ACGAN architecture. The generator G takes in z, the random noise, and c, the class label. The output Q is the discriminator's probability distribution of the class labels, and y is the probability that the input to the discriminator is real.

The selection of ACGAN for this task is optimal since it allows for the use of a diverse array of class labels for training the model. If a Deep Convolutional GAN (DCGAN) or a WaveGAN (Donahue et al., 2018) were used, only the COVID-19 audio segment could be used for training the model; however, more class labels can be used to train the ACGAN since the generator is given the class label of the desired data.

## 4. Data Collection and Preprocessing

The selected dataset for cough augmentation is the coughvid dataset (Orlandic et al., 2021), a large COVID-19 audio dataset. The coughvid dataset consists of a total of 34,434 cough samples of different patients with diverse ages, genders, and geographical locations, 1155 of which belong to a COVID-19-positive patient. Furthermore, as an added layer of validation, the authors had four experts evaluate a fraction of the dataset. In total, 2,800 samples were evaluated by an expert. The audio samples were noted to have multiple cough instances, so each sample was segmented into multiple segments of uniform length.

*4.1 Data analysis*

This phase involves examining the distribution of the data and pinpointing any inherent biases within the cough samples. Insights garnered from this preliminary data investigation will guide the selection and preprocessing stage of the pipeline.

**Data Biases**: The analysis shown in Fig. 2 reveals that numerous participants, likely due to the nature of voluntary participation, opted to leave many optional fields blank. An initial review of the data indicates a pronounced gender imbalance, with male participants outnumbering females by a factor of two. This gender imbalance is not expected to negatively impact training as COVID-19 symptoms – including coughs - have no gender-based differences.

Additionally, when examining patient status distribution, there's a notable deficiency in COVID-related samples, which are limited to approximately 1,000, in contrast to other statuses that span tens of thousands. Further, the data suggests a predominance of participants without respiratory conditions, and similarly, most reported an absence of fever or muscle pain.

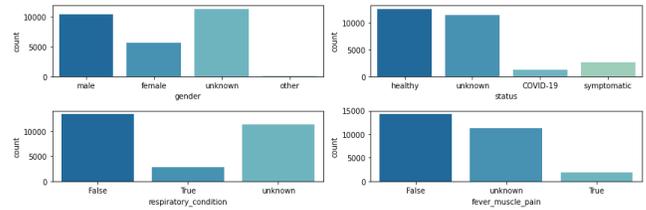

**Figure 2**. Data Biases analysis.

**Quality examination:** The evaluation of the audio samples' quality is conducted through the analysis of two specific attributes: 'cough_detected' and 'SNR'. 'cough_detected' signifies the likelihood that a cough is present within the audio clip. An investigation into this attribute's distribution reveals a significant probability of the inclusion of non-cough sounds, as shown in Fig3. This is addressed in the data preparation phase.

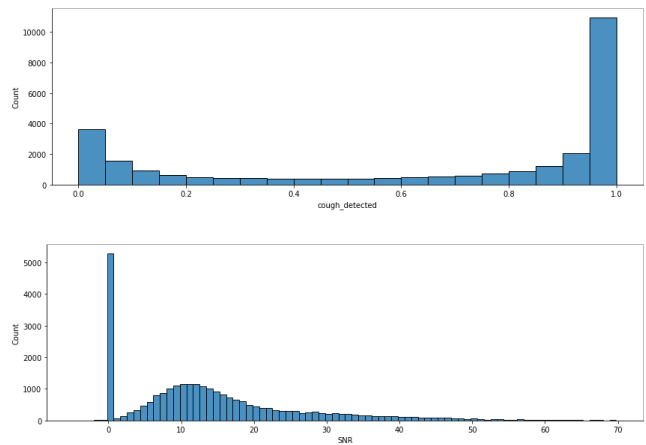

**Figure 3**. Quality examination.

**Cough diagnosis and severity:** physicians conducted an examination of the cough recordings, offering a diagnostic assessment for each, coupled with an evaluation of the cough's severity, as illustrated in Figure 4. Despite the initial categorization of data samples into healthy, symptomatic, COVID-19 infected, or unknown groups, the doctors classified the coughs into categories of healthy, lower respiratory infection, COVID-19 infected with obstructive pulmonary disease, upper respiratory infection, or unknown. The general severity of the conditions was deemed mild, and as clarified in Fig. 5, the correlation between the severity and the diagnostic categories revealed no irregularities.

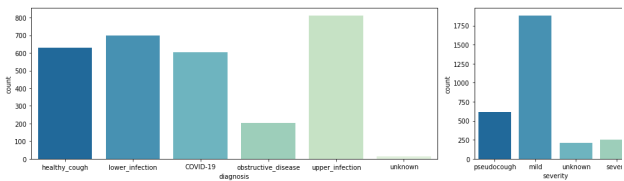

**Figure 4**. Left. Cough diagnosis. Right. Cough severity.

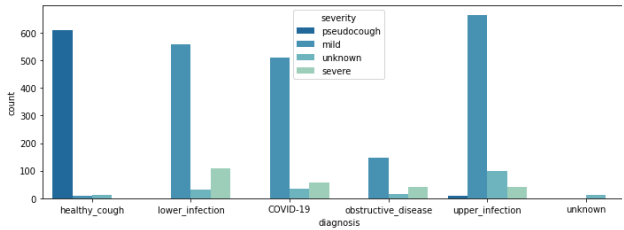

**Figure 5**. Physicians' diagnosis with respect to the severity of the patient's case.

### 4.2 Data preparation

Following the initial data exploration and necessary cleaning, the subsequent action involves the preprocessing of the audio data. Figure 6 presents a randomly chosen example from each of the three audio categories: healthy, symptomatic, and COVID-19. Upon reviewing the samples, it is observed that the audio recordings exhibit variations in their durations and are often interspersed with significant amounts of silence or ambient noise. Therefore, the preprocessing phase is tasked with more than just normalizing and rendering the audio suitable for analytical models; it must also ensure the audio clips are of uniform length and effectively address issues related to noise and silent periods.

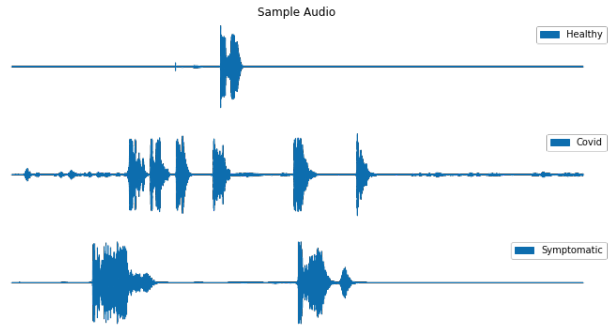

**Figure 6**. Randomly selected sample from each of the three classes of audio.

All cough recordings with a missing status_SSL label or a cough_detected value below 0.7 have dropped out.

The cough audio files remaining in the dataset were first preprocessed by normalizing them to the range of -1 to 1, a butter low-pass filter was applied, and the signals were down sampled to 12 KHz.

The processed cough audios were then segmented into smaller segments implemented by Orlandic et al. The segmentation algorithm is based on a hysteresis comparator with two thresholds, high and low. When the signal crosses the high threshold, a new segment begins and remains active until the audio signal drops below the low threshold.

The high and low thresholds are set based on the root mean square (RMS) of the full signal. The high threshold was set at twice the RMS, and the low threshold was set at 0.1 times the RMS. Further, all detected segments are padded with 0.1 seconds (0.05 at the beginning and 0.05 at the end) to ensure the minimum segment length is 0.1 seconds. The individual segments were converted to decibel scale Mel Spectrograms using the librosa Python audio library. The Mel spectrograms are generated with a 2048 FFT window, 512 hop length, and a Hanning window.

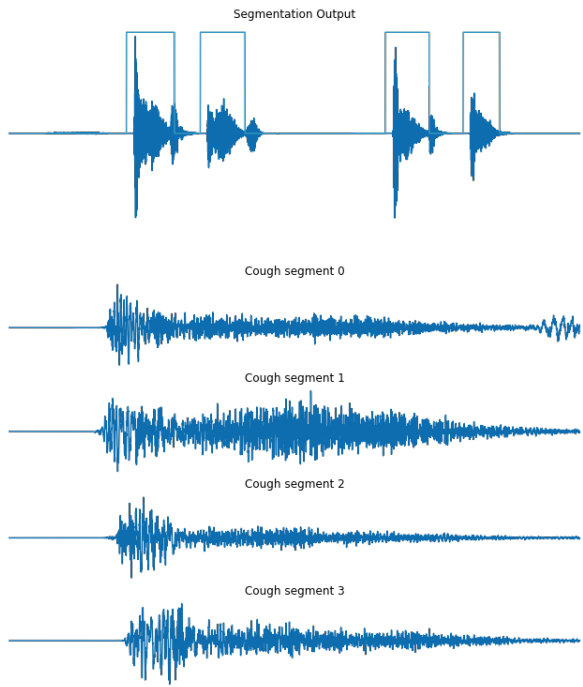

**Figure 7**. Generated cough segments.

*4.3 Feature extraction*

After the data preprocessing phase, the next step involves extracting features from the audio data. In the realm of deep learning models tailored for audio-related tasks, it's a standard approach to visually depict the audio's features and utilize these visual representations for the task at hand. Interestingly, there exists a variety of algorithms dedicated to the representation of audio features, and choosing the most suitable algorithm is crucial for the success of the audio task. The ensuing subsections delve into various feature extraction algorithms, with each section presenting outcomes based on analyses conducted on a randomly chosen audio sample, as illustrated in Fig. 8 (a).

**Spectrogram**: it visually portrays the intensity, or "volume," of a signal across time, spanning various frequencies found within a waveform. This allows for the observation of energy distribution across frequencies, for instance, comparing the energy at 2 Hz against that at 10 Hz, as well as tracking the fluctuations in energy levels as time progresses. Both Short-Time Fourier Transform (STFT) and Mel-spectrogram are used to generate this feature as clarified in Fig 8 (b and c).

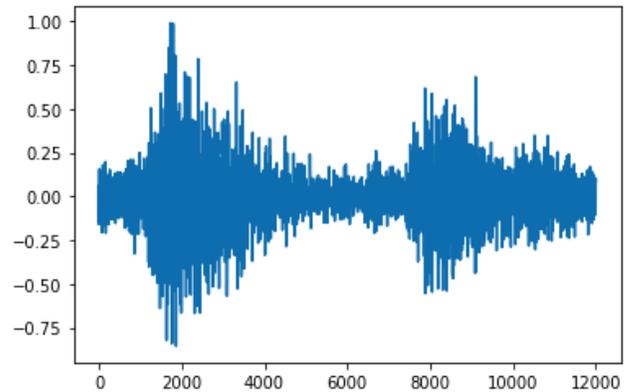

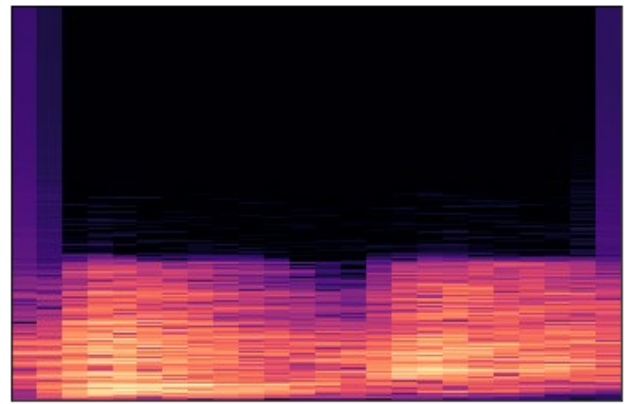

(b) SIFT

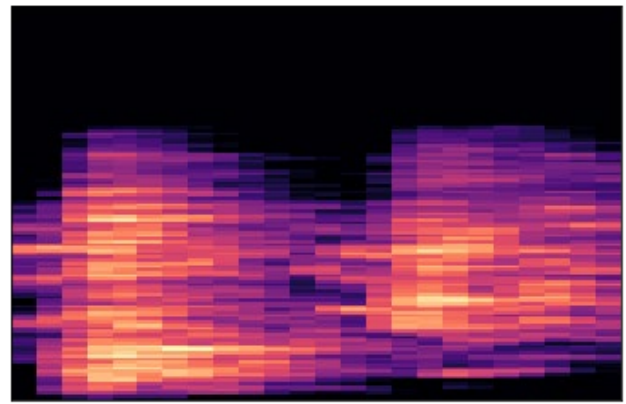

(c) mel-spectrogram

**Figure 8**. Generated cough segments.

# 5. ACGAN Architecture

## 5.1 Discriminator

The discriminator takes in a spectrogram of the uniform shape and then processes it through five convolutional layers enhanced with batch normalization, LeakyReLU -alpha 0.2, and 50% dropout. 32, 64, 128, 256, and 512 convolutional filters were used, respectively. The kernel size is 3x3 for all convolutions, and only the first convolution used a 1x1 stride, while the rest used a 2x2 stride. Then, the output is flattened and passed to two different layers, one for validation and one for labeling. The validity layer determines whether the given input is real or fake and uses a sigmoid activation function, while the label layer generates a probability distribution representing the likelihood of the given image belonging to a specific class and uses the sigmoid activation function.

To train the model in a way that would optimize its ability to classify input and validate it, two loss functions are used. First, the model is optimized using the Adam optimizer with a small learning rate -0.0002- which was proven to provide stable training. As for the loss functions, the binary cross-entropy was selected. The binary cross-entropy optimizes the binary task of validating the realness of the input as well as measuring the binary classification loss of the model.

## 5.2 Generator

While the ACGAN paper defined that the generator model should take a vector input that is a concatenation of the latent space and the class label, modern approaches separate the two into branches: the noise branch and the label branch. The label branch is perceived as an additional feature map which is achieved by using a learned embedding with an arbitrary number of dimensions, 50 in this case. The output of the embedding layer can then be passed to a fully connected layer with a linear activation function to generate a 7×7 feature map.
The noise from the defined latent space of 512 is interpreted by a fully connected layer with a ReLU activation function to create 1024×16×3 feature maps, each feature map representing a low-resolution version of the expected output. The noise and label are then concatenated channel-wise before being passed to the model.

Through a series of 2D transpose convolutional layers, the 16×3 feature map is up-sampled by an iterative factor of two until the desired output shape of 128×24 is achieved. Relu activation functions are used in place of leakyRelu following the ACGAN paper. Finally, using a tanh activation function, the output of the generator is a single feature map with pixel values between [-1,1].

For each batch, the image batch and respective labels of those spectrograms are generated then the generator model is tasked with generating spectrograms for different class labels. The discriminator is training on the real images and their labels and its loss during that training is labeled as real loss representing the class and validation loss on real data. Then the model is trained on the generated samples where the loss measured is labeled fake loss. Both losses are recorded to be displayed after each epoch.

Then after the discriminator's weight has been updated with backpropagation the composite model is trained with generated noise and randomly selected labels. Since the composite model consists of both the generator and discriminator the noise and class labels will be converted into generated images that the discriminator will then label. Since the two models are connected the loss of the discriminator will backpropagate to the generator without altering the weights of the discriminator. Since the discriminator was updated before the generator the loss will improve the generator's performance.

The generator network is optimized using the loss of the discriminator. This is achieved by creating a composite model comprising both the generator and discriminator such that the weights of the discriminator are not updated during the backpropagation of the composite model. What this achieves is an indirect way to optimize the generator. Since the composite model uses the discriminator loss, it will be compiled in the same way as the discriminator.

## 6. Training Methodology

A standard train-validation-test split is made on the Mel-spectrograms for each segment. The split was done in a ratio of 80%-10%-10% respectively, to cope with the limited data availability. The initial baseline CNN was trained on the task of Healthy vs. COVID-19 Mel classification on the spectrograms, with the training hyperparameters listed in Table 1.

**Table 1.** Hyperparameters for the baseline CNN training

| Hyperparameter | Value |
| --- | --- |
| Learning Rate | 0.002 |
| Epochs | 200 |
| Batch Size | 256 |
| Beta 1 | 0.9 |
| Beta 2 | 0.999 |
| Weight Decay | 0.01 |

Following up, the ACGAN was trained with the same training split to train the baseline classifier. This is important given that the synthetic samples generated by the ACGAN are meant to be seen as data augmentations of the original training set. The training hyperparameters for each of the generator and discriminator are outlined in Tables 2 and 3, respectively.

**Table 2.** Hyperparameter values for the training of the ACGAN generator

| Hyperparameter | Value |
| --- | --- |
| Latent Dimension | 512 |
| Learning Rate | 0.0002 |
| Beta 1 | 0.5 |
| Beta 2 | 0.999 |
| Epochs | 1000 |

**Table 3.** Hyperparameter values for the training of the ACGAN discriminator

| Hyperparameter | Value |
| --- | --- |
| Learning Rate | 0.002 |
| Beta 1 | 0.5 |
| Epochs | 1000 |
| Gaussian Noise Mean | 0 |
| Gaussian Noise Initial Variance | 0.1 |

While passing both real images into the generator during training, Gaussian noise has been added. This approach is inspired by the paper on Instance Noise. It's claimed that by adding random noise to both the real images and fake images, their underlying distribution can overlap more. The noise increases the difficulty of the learning task, but also makes it more stable through the overlap. This leads to improved convergence as the log-likelihood ratios between the real and fake distributions are well defined.

The random noise was added pixel-wise with a mean of 0 and starting variance of 0.1. The variance value was annealed through training and decreased linearly with each epoch until a final variance of 0 was reached on the final epoch. Soft labels were also used in the discriminator when evaluating the losses for both real and synthetic images. Labels with value 1 were replaced with a random value sampled uniformly from the interval 0.8 to 1. Similarly, labels with value 0 were replaced with a random value sampled uniformly from the interval 0 to 0.2. The goal of the instance noise and soft labeling techniques was to provide an additional challenge to the discriminator, as it has been observed that it can otherwise overpower the generator, and the discriminator loss quickly falls to 0. The latent noise for the generator was a 512-dimensional vector which sampled elementwise from a Gaussian distribution with mean 0 and variance 1. Once the ACGAN training was completed, 200 synthetic images, corresponding to 25% of the initial training set size, were conditionally generated from each class and used to augment the training set. A new baseline CNN model with the same initial architecture was retrained on the augmented dataset containing the original real samples along with the synthetic images.

## 7. Results

### 7.1 Establishing a baseline classifier

Initially, a convolutional network is trained to establish a baseline classification accuracy of the spectrograms into their respective classes. The purpose of the baseline accuracy is to establish whether an improvement in the classification accuracy can be made when the ACGAN synthetic data is added to the training dataset. This is done by randomly sampling an equal number of *healthy, symptomatic, and COVID-19* cough spectrograms. It should be noted that these labels are self-reported according to the dataset maintainers and do not require a PCR test to confirm the label.

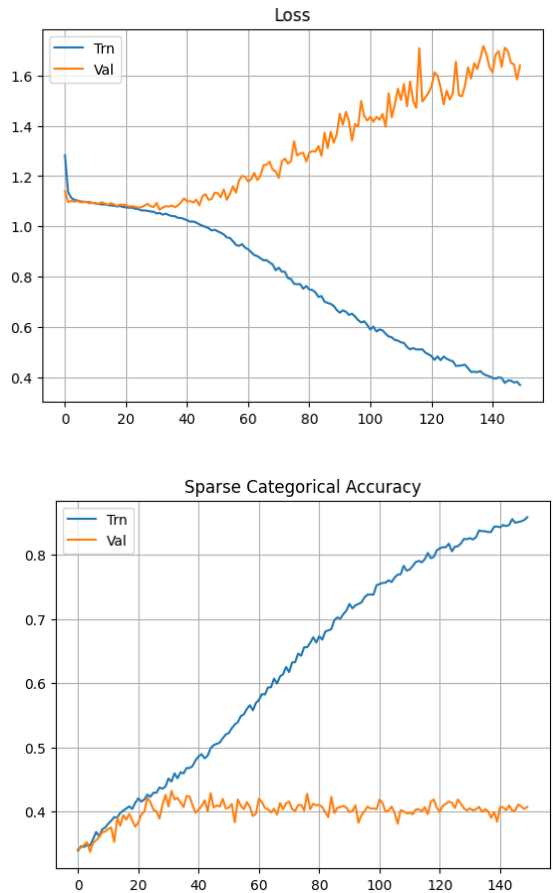

**Figure 9**. Model training.

As seen in the figure above, while the model continuously increases in training accuracy as the training process proceeds, the validation accuracy struggles to increase beyond 41%. This indicates the overfitting of the model on the training data. Instead, a pretrained ResNet50 is fine-tuned on the spectrograms to observe whether the training process improves. The backbone ResNet50 weights are frozen, and the backbone output is flattened and passed through a LeakyReLU layer, which is finally into a SoftMax layer with three output neurons. An Adam optimizer with a 0.001 learning rate is used and a sparse categorical cross-entropy loss.

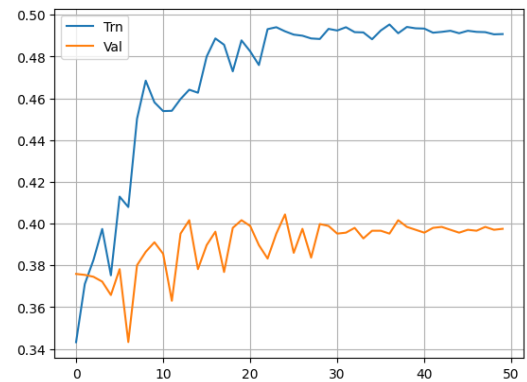

**Figure 10**. Fine-tuning ResNet50 on COVID spectrograms.

Instead, a network based on 1D convolutions was constructed and trained in the same manner using categorical cross-entropy loss and Adam optimizer.

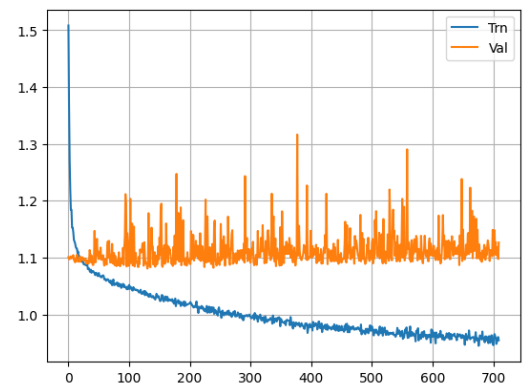

**Figure 11**. Cross entropy loss using Conv1D network.

The training process is run for much longer, at 700 epochs, but the outcome, while being much noisier, is still the same, with the model overfitting on the training dataset while the validation accuracy score oscillates about the 41% value.

After switching to the revised dataset by Orlandic (2023) containing the self-supervised data results, a new CNN classifier was trained with improved results. The architecture of the classifier matches the discriminator architecture without the validation layer.

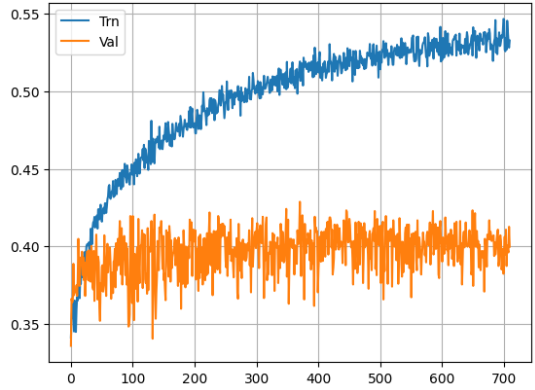

**Figure 12**. Accuracy using Conv1D network.

While the classification rate improved significantly on the validation, it reached a maximum value of 72% on the validation set while the classifier began to overfit the training data.

### 7.2 ACGAN Training

This provides a working value to improve upon. The ACGAN is trained following the procedure mentioned in the discussion above.

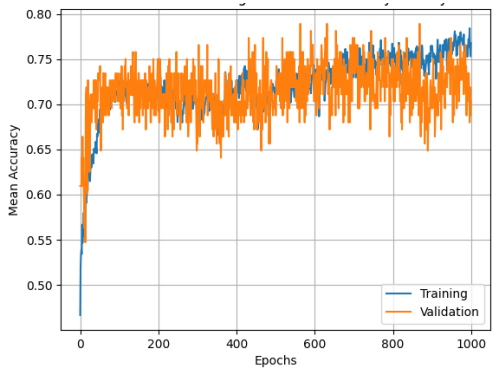

**Figure 13**. Discriminator categorical real accuracy history.

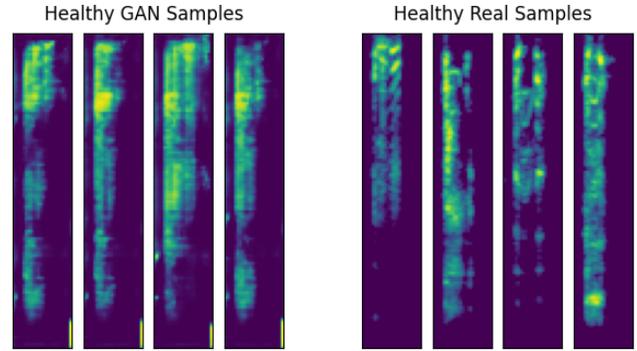

**Figure 14**. Healthy GAN vs Real Healthy Samples.

The plots to the left in both figures above display two sets of synthetic Mel spectrograms generated by the ACGAN. The plots in the upper figure to the right display real COVID-19 Mel spectrograms. The GAN samples include some repeating artifacts, such as the blobs centered in the middle of the figures, as well as decreased energy in the lower frequencies of the spectrograms. On the other hand, the real Mel spectrograms have a characteristic columnar shape that is also lacking in the generated samples.

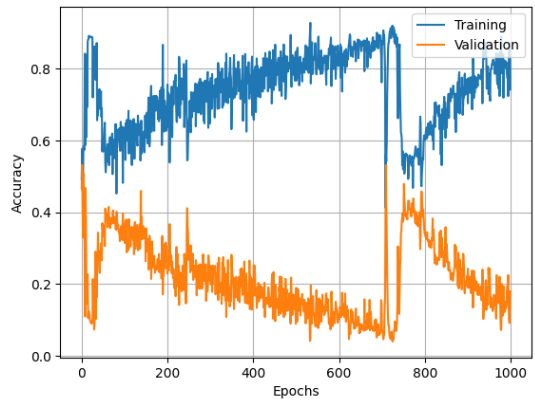

**Figure 15**. Mean discriminator adversarial probability.

When the ACGAN training begins, the discriminator weights are randomly initialized, and it assigns equal probabilities to both real and synthetic samples. As the training progresses, the discriminator's assigned probabilities to the respective classes, real and synthetic, tend towards their target values, 1 and 0, respectively, with a momentary trend towards 0.5 around the 50-epoch mark. This is, however, not a sign of convergence, given that the generated samples at that interval were inspected and observed to be of lower quality in comparison to the real samples. This indicates that the discriminator was being temporarily overpowered by the generator.

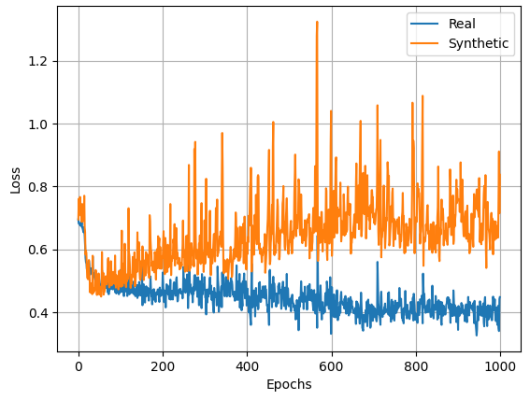

**Figure 16**. Mean discriminator vs generator loss.

The diagram highlights the training dynamics of both the discriminator and generator. While the discriminator eventually dominates, the samples generated by the generator remain visually plausible.

Lastly, 1000 novel samples from the "COVID-19" and "Healthy" class labels are conditionally generated using the ACGAN. These augmented samples are mixed into the training dataset to train a new classifier CNN once again.

A CNN architecture identical to the baseline CNN architecture is used, except for the output layer, where only one sigmoidal output node is used for the classification task. It should be noted that throughout the entire training procedure, from establishing the baseline accuracy to training the ACGAN, an independent testing dataset is held out to evaluate the performance of the new CNN trained on the augmented training set.

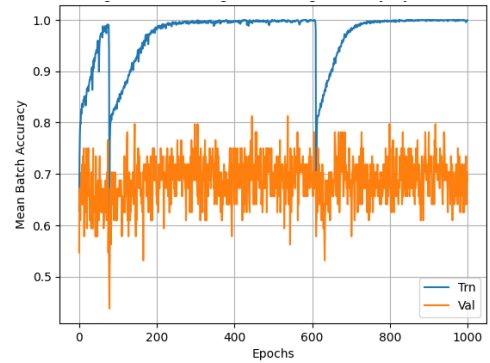

**Figure 17**. training CNN on the augmented dataset.

The figure above demonstrates the accuracy dynamics during training of the CNN on the augmented dataset. The CNN training is quick to overfit on the augmented set while occasionally dropping back down. However, at the end of training, the trained CNN was evaluated on the testing set once more, with the classification accuracy increasing to 75%.

## 8. Discussion

We observe only a modest improvement in the classification accuracy from 72% in the baseline model to 75% in the retrained model.

When the conditionally generated synthetic Mel spectrograms are converted back into audio format, the audible quality of both classes was moderate in terms of resemblance to a cough sound. We cannot state whether the COVID-19 samples in particular exhibited notable traces of the disease, given our lack of medical training; however, they exhibited a symptomatic tone relative to the healthy synthetic samples. Nonetheless, we identify several potential challenges, such as the instability in training the GANs, leading to mode collapse towards the majority class, given that the healthy class yielded more diverse and higher quality synthetic audio from our perspective. The stability issue of training GANs is well known in the literature. Our training techniques leveraged several well-known techniques to help stabilize the training, as mentioned in the training methodology.

We also wonder if the cough segmentation strategy may have been too sensitive and prone to splitting single coughs, with intermediate breaks, into two independent segments. The cough physiology has three phases characterized as inspiratory, compressive, and expiratory (Lee et al.). The compressive phase is a short intermediate silent phase that gets detected by the segmentation strategy as a cutoff point that splits the cough into two separate segments. This doesn't happen for all the coughs in the dataset but rather depends on how short and silent the compressive phase sounds in the cough audio. A suggested fix for this would be to increase the allowed time for the signal to be below the proposed threshold and increase the minimum audio signal length.

Lastly, the volume of data labelled as COVID-19 was simply too little in addition to having been labelled by an SSL algorithm, which poses a challenge for either of the classification model or the generator model to learn effective representations of the COVID-19 class. The scarcity of data carried over as well into the Healthy class, as we tried to balance the number of samples used per class. Furthermore, the moderate inter-rater reliability as well as the self-reported statuses of the coughs. In the initial classifiers built around the `status` column values containing the self-reported statuses without a PCR test confirmation, modest accuracy values were achieved by the classifier in distinguishing between healthy, symptomatic, and COVID-19. This issue was identified by Orlandic et al., who decided to leverage the expert-annotated data to train a self-supervised learning model to provide new labels for a smaller subset of the data. The new labels were limited to healthy and COVID-19. Using these new status_SSL labels improved the performance of our baseline classifier to 72%. However, it came at the cost of a reduced dataset.

There is existing work on COVID-19 based on both classification and detection. The COVID-19 classification work looks at classifying audio samples either by extracting audio DSP features (as done in Orlandic et al.) or by leveraging CNNs to operate on the Mel spectrograms (Hamdi et al.), while contending with the issue of class imbalance. Other works take up the task of radiological data synthesis by leveraging GANs to generate CET scans (Jiang et al.) of COVID-19 positive patients to make up for the data imbalance. Our work provides a novel contribution by attempting to tackle the audio data imbalance by synthesizing audio samples, which can serve as a means of data augmentation to improve COVID-19 audio-based classifier accuracy.

## 9. Conclusions

Based on the findings, this study concludes that the application of an Auxiliary Classifier GAN (ACGAN) to synthesize COVID-19 cough audio demonstrates the potential to alleviate data scarcity issues in COVID-19 audio-based classification. By generating synthetic Mel spectrograms of both healthy and COVID-19 coughs, the ACGAN-supported dataset augmentation led to a modest improvement in classifier accuracy, increasing from an initial 72% to 75%. This improvement highlights ACGAN's potential to provide diverse synthetic data that enhances model training in other constrained data settings, for example, either due to novel diseases or when dealing with issues around data scarcity due to the challenges of patient privacy, sensitivity, and the costs of collecting real data. Likewise, a key part of the ACGAN has been training an effective discriminator, which must, in part, act as a classifier. This work contributes to demonstrating the potential of using convolutional networks in audio-based diagnosis.

Potential avenues for future research include exploring more recent generative modelling techniques that are transformer or diffusion-based to increase the quality of the generated spectrograms, as well as exploring deep self-supervised learning techniques to provide labels to a larger portion of the dataset, which can bootstrap the training process.

However, the study also underscores the inherent challenges in synthesizing audio signals that authentically capture the nuanced characteristics of COVID-19 coughs. These include challenges with segmentation methods, the presence of artifacts in synthetic samples, and limitations in current classification metrics. Some avenues for future exploration include adding alternative datasets to the mix, which may improve the generalization ability of the built classifier and support the ACGAN in generating more novel outputs. Furthermore, focusing on alternative cough segmentation techniques that better preserve the entire cough segment may be required.

A more critical issue was the presence of conflicting and mislabeled samples within the COVID-19 positive and negative categories. For instance, labels often relied on self-reported data, where participants may have claimed to be COVID-19 positive without a confirmed RT-PCR test. Furthermore, audio recordings were sometimes inaccurately categorized, reflecting either the participant's misunderstanding of their health status or inconsistencies in expert review during data annotation. As a result, the datasets included instances where COVID-19 cough samples might have been misclassified as healthy or symptomatic, leading to noisy training data and confusing signals for the machine learning models.

This mislabeling affected the overall quality of the synthesized samples and impaired the classifier's ability to accurately distinguish between COVID-19 positive and healthy coughs. Even after synthesizing additional data using ACGAN, these inconsistencies impacted the model's performance, which introduced artifacts and limited the model's classification accuracy. Addressing these issues will require more robust data collection methods, better-defined labeling criteria, and quality control mechanisms, such as using confirmed RT-PCR results to verify labels or implementing more reliable semi-supervised labeling algorithms for noisy data. Until the development of methods and equipment to meet such requirements, data scarcity will remain a critical factor that even impacts data augmentation tasks.